# Interface Bonding of $Zr_{1-x}Al_xN$ Nanocomposites Investigated by X-ray Spectroscopies and First Principles Calculations

Martin Magnuson, Weine Olovsson, Naureen Ghafoor, Magnus Odén and Lars Hultman

*Department of Physics, Chemistry and Biology (IFM) Linköping University, Sweden*

2020-03-18

## Abstract
The electronic structure, chemical bonding and interface component in ZrN-AlN nanocomposites formed by phase separation during thin film deposition of metastable $Zr_{1-x}Al_xN$ ($x$=0.0, 0.12, 0.26, 0.40) is investigated by resonant inelastic X-ray scattering/X-ray emission and X-ray absorption spectroscopy and compared to first-principles calculations including transitions between orbital angular momentum final states. The experimental spectra are compared with different interface-slab model systems using first-principle all electron full-potential calculations where the core states are treated fully relativistic. As shown in this work, the bulk sensitivity and element selectivity of X-ray spectroscopy enables to probe the symmetry and orbital directions at interfaces between cubic and hexagonal crystals. We show how the electronic structure develop from local octahedral bond symmetry of cubic ZrN that distorts for increasing Al content into more complex bonding. This results in three different kinds of bonding originating from semi-coherent interfaces with segregated ZrN and lamellar AlN nanocrystalline precipitates. An increasing chemical shift and charge transfer between the elements takes place with increasing Al content and affects the bond strength and increases resistivity.

## 1. Introduction
Transition metal nitride compounds in the form of thin films are technologically important as thermally and mechanically stable ceramics in applications such as cutting tools, bearings, diffusion barriers, as well as contact layers in microelectronics and piezo-electric, thermoelectric, and decorative coatings [1]. Alloying the stable binary transition metal nitrides TiN, CrN, ScN, and ZrN with Al results in metastable pseudo-binary ternary alloys such as $Ti_{1-x}Al_xN$, $Cr_{1-x}Al_xN$, $Sc_{1-x}Al_xN$, and $Zr_{1-x}Al_xN$, respectively. These alloys display a large miscibility gap and limited solubilities at temperatures below 1000º C. However, solid solutions of these alloys can be synthesized by physical vapor deposition techniques such as magnetron sputtering and arc-deposition which effectively quench bulk diffusion against extended phase separation. Such metastable solid solutions will phase separate when heated, which is attractive in many cases because a microstructure consisting of nm-sized domains are formed that positively affect the functional properties of the material, like for hardening. The transformation may also be initiated and controlled already during the film syntheses by choosing proper deposition parameters yielding a wide range of achievable microstructures, including different types of nanocomposites. Common to these alloys is that the addition of Al and formation of internal interfaces alter the atomic bonding. It is noteworthy that despite the rather large number of studies on these alloys, the effects of Al-alloying and internal interfaces on the electronic structure have been overlooked to date.

Among these metastable alloys, the $Ti_{1-x}Al_xN$ system is the most well studied and it has a large miscibility gap. Thus, $Ti_{1-x}Al_xN$ alloys grown as solid solutions have a large driving force for the formation of a two-phase system of TiN and AlN [2]. When heated, this results in spinodal decomposition into TiN and AlN-rich domains resulting in age hardening [3]. Alternatively, a phase-separated microstructure can be achieved by increasing the growth temperature and in this case the transformation occurs during growth at the growth front [4]. Similarly, $Sc_{1-x}Al_xN$ can be grown both as solid solutions and nanocomposites, which impact the piezoelectric properties [5] [6] [7]. The miscibility gap of $Cr_{1-x}Al_xN$ is considerably smaller than $Ti_{1-x}Al_xN$ and the solid solution state is more easily retained. Alloying $Cr_{1-x}Al_xN$ with Ti offers possibilities to tune both decomposition path and oxidation resistance [8] [9]. In the case of $Zr_{1-x}Al_xN$ alloys, which have an even larger miscibility gap than $Ti_{1-x}Al_xN$ solid solutions, these are difficult to stabilize and instead two-phase systems tend to form already during deposition. The strong driving force for decomposition results in a large variety of microstructures that primarily are determined by the growth temperature and Al content [10] [11]. Common to all of these alloys is that the nature of the interatomic bonding and electronic structure are affected by the alloy composition and that they change during decomposition. Decomposition generally also results in formation of a high density of internal interfaces so-called nano-labyrinths, where the electronic structure is further altered [12], which in most cases is ignored.

In this study we have chosen $Zr_{1-x}Al_xN$ for alloy nitride materials model system to probe changes in electron structure and chemical bonds as functions of alloy composition and formation of internal interfaces. The growth and resulting microstructures of the studied $Zr_{1-x}Al_xN$ alloys have been reported in detail elsewhere, primary using X-ray diffractometry (XRD) and transmission electron microscopy (TEM) [10] [11]. At high Al contents a two-phase nanocomposite is formed consisting of cubic ZrN (space group Fm3m, B1, 225) and hexagonal AlN (wurtzite, space group P63mc, B4, 186) with chemically sharp semi-coherent interphases. At low Al contents, the interfaces are crystallographically coherent although chemically more diffuse since both Zr-rich and Al-rich domains have the same cubic (B1) crystal symmetry. All these microstructures result in high hardness and the highest hardness is found in the range of $0.2 < x < 0.5$, *i.e.* where the interfaces are semi-coherent.

The sensitivity of X-ray spectroscopy enables us to probe the orbital directional occupations to investigate the bonding structure at the internal interfaces that determines the bond characteristics (symmetry and orbital directions at interfaces between cubic and hexagonal crystals) and local chemistry that affects conductivity and hardness [13]. The cubic-AlN phase that may form has a bandgap of about 6 eV [14] while it is 3 eV for ZrN [15]. Theoretically, it has been found that for increasing $x$ in $Zr_xAl_{1-x}N$, the Fermi level ($E_F$) is shifted into the conduction band [16] [17]. A semiconductor-to-metal transition has therefore been predicted for $x < 0.25$. These theoretically predicted variations in metallicity and bond characteristics depend on the local chemistry and emphasize the importance of understanding the bond characteristics at the internal interfaces through experiments.

Previous spectroscopic investigations of ZrAlN thin films have utilized surface-sensitive X-ray photoemission spectroscopy (XPS) where electrons were detected at the Zr *3d* levels [18] [19]. However, XPS is not ideal for probing the electronic structure and chemical bonding in the bulk of the material's interior and to distinguish the bonding structure between semi-alloys, compounds and nanocomposites [13] [20]. These types of materials require deep probing techniques, which X-ray spectroscopy that detects photons can provide.

In this paper, we employ bulk-sensitive and element-selective soft X-ray absorption (XAS) and X-ray emission spectroscopy (XES) to investigate the local short-range order of the bond structure at the internal interfaces in $Zr_{1-x}Al_xN$ films with increasing Al content in comparison to modeled structures using density-functional theory (DFT) calculations of ordered alloys and interfaces. In particular, the valence band shifts and charge-transfer between nm-sized AlN domains and the surrounding Zr-based matrix reveal the details in the electronic structure. Such knowledge paves the way for coating improvements, through for example alloying to achieve improved wear protections at cutting tools edges or enhanced piezoelectric response in energy harvesting applications [21].

## 2 Experimental details

### 2.1 Thin film synthesis
The $Zr_{1-x}Al_xN$ thin films were grown on MgO(001) substrates to a thickness of 250 nm by reactive magnetron sputtering using a deposition system with a base pressure of $1*10^{-6}$ Torr. Depositions were conducted in an Ar-$N_2$ gas mixture at $4.5*10^{-3}$ Torr by co-sputtering from two 75 mm elemental targets, Zr (99.9%) and Al (99.999%) mounted off-axis at an angle of 25° with respect to the substrate surface normal. The substrates were kept at 700 °C and rotated by 60 rpm. The composition of the films was varied by changing the applied power to the two metal sources such that the aluminum content was increased from $x$=0, 0.12, 0.26 to 0.40 as described elsewhere [10]. In addition, a pure Zr film was also grown on a MgO(001) substrate. The Zr and ZrN films are used as reference.

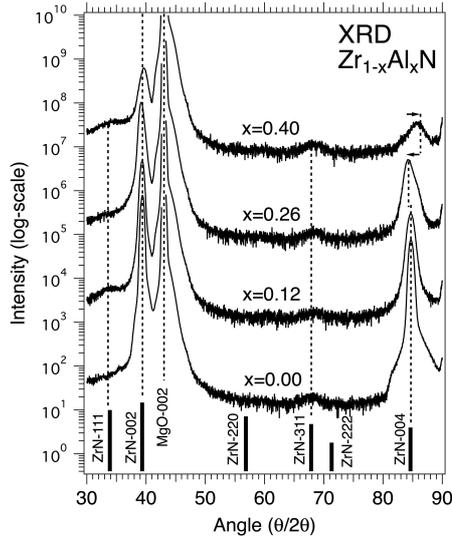

**Figure 1:** X-ray diffractograms of the $Zr_{1-x}Al_xN$ thin films grown on MgO(001) substrates. The vertical bars indicate c-ZrN reflections.

The phase contents of the films were characterized by X-ray diffraction (XRD) θ/2θ scans in a Philips powder diffractometer, using Cu $K\alpha$ radiation at 40 kV and 40 mA. The primary beam was conditioned using 2x2 mm² crossed-slits and in the secondary beam path a 0.27° parallel plate collimator was used together with a flat graphite crystal monochromator. A proportional detector was used for the data acquisition.

Figure 1 shows θ/2θ X-ray diffractograms (XRD) data of $Zr_{1-x}Al_xN$ samples ($x$=0.12, 0.26, 0.40) in comparison to c-ZrN ($x$=0.00). For c-ZrN [22] with space Group Fm3m (225), the diffractograms are dominated by an intense ZrN-002 peak at 2θ=39.404°, a MgO-002 substrate peak at 43.0° and a ZrN-004 peak at 2θ=84.751°. In addition, features with low intensity are observed around 34° and 68°, related to c-ZrN-111 and c-ZrN-311 planes. Note that the diffraction from c-AlN and w-AlN is negligible in comparison to c-ZrN. From the c-ZrN-002 and c-ZrN-004 peak positions, the lattice parameter of the c-ZrN film was determined to $a$=4.5698 Å in good agreement with the literature value of bulk c-ZrN (4.5675 Å) [22]. As the Al content is increased to $x$=0.12, the intensities of the 002 and 004 peaks decrease and there is a small expansion of the c-ZrN lattice parameter to $a$=4.5732 Å, as observed by the c-ZrN peak shifts towards slightly lower angle. The reduced intensity of the 002 and 004 peaks and the lattice expansion is more obvious for $x$=0.26, where $a$=4.5849 Å. For $x$=0.40, the intensities of the 002 and 004 peaks become even lower and the peaks significantly broaden while there is a contraction of the lattice to $a$=4.5412 Å, as observed by the ZrN peak shifts to higher angles. The large broadening of the peaks is an indication of reduced c-ZrN grain sizes (nano-crystallites or a smaller average domain size) as the peak widths are inversely proportional to the crystallite size [23]. Previous studies indicated that starting from $x$=0.2, the ZrAlN system has a two-phase structure [24] [11]. With increasing Al content, the $Zr_{1-x}Al_xN$ solid solution ($x$=0, 0.12, 0.26) gradually transforms to a nanocomposite ($x$=0.4-0.55) with a characteristic two-phase laminated structure described in ref. [24].

## 2.2 X-ray emission and absorption measurements

The sample's electronic structure and orbital occupation were investigated using bulk-sensitive and element-specific XAS and XES to probe the unoccupied and occupied bands of the different elements. By tuning the energy of linearly polarized X-rays to the specific core levels of Zr, N, and Al spectra of the occupied and unoccupied electronic orbitals were obtained.

The XAS and XES spectra were measured at normal and 15 degrees incidence angle at 300 K and ~ $1·10^{-8}$ Torr at the undulator beamline I511-3 [26] at the MAX II ring of the MAX IV Laboratory, Lund University, Sweden). The XAS energy resolution at the N $1s$ edge of the beamline monochromator was 0.2. The XES spectra were recorded with spectrometer resolutions of 0.4, 0.5, and 0.1 eV for the Zr $M_{4,5}$, N $K$ and Al $2p$ emission, respectively. The XAS spectra were measured in bulk-sensitive total fluorescence yield (TFY) and normalized by the step edge below and far above the absorption thresholds. The XES spectra were measured at 20 degrees incidence angles. For comparison of the spectral profiles, the measured XES data were normalized to unity and were plotted on a common photon energy scale (top) and relative to the Fermi level $E_F$ (bottom).

# 3 Computational details

## 3.1 First-principles calculations

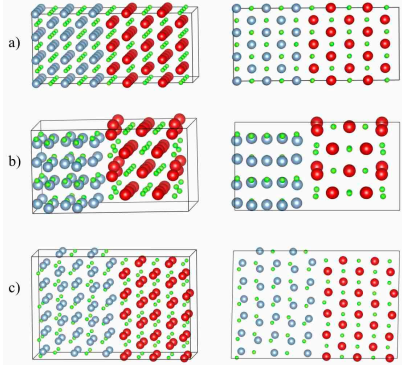

**Figure 2:** Model structures for three different interface slabs: a) c-ZrN//c-AlN-100, b) c-ZrN//w-AlN-110 and c) asymmetric c-ZrN//w*-AlN-001.

All *ab initio* calculations were performed within Density Functional Theory (DFT) [27], utilizing the Generalized Gradient Approximation (GGA) exchange-correlation functional according to Perdew *et al.* [28]. In a first step, geometrical relaxations were considered by applying the Vienna ab initio simulation package (VASP) [29] [30], which is based on the Projector Augmented Wave (PAW) [31] method. Finally, X-ray spectroscopy calculations were handled by the all-electron full potential linear augmented plane wave (FPLAPW) WIEN2k software package [32], using the augmented plane wave plus local orbital (APW+lo) basis set.

Figure 2 shows model structures for three different interfaces: a) cubic/cubic c-ZrN//c-AlN-100, b) symmetric cubic/hexagonal c-ZrN//w-AlN-110 and c) asymmetric cubic/hexagonal c-ZrN//w*-AlN-001. The structures are abbreviated as c/c, c/w and c/w*. The model structures were chosen with the fully cubic interface as a reference system while the two different cubic/wurtzite interfaces were selected based on observations by transmission electron microscope (TEM) [10]. For the three different interface systems, the Al concentration is x=0.5 for c/c and c/w and *x*=0.55 for c/w*. The modeled structures also include cubic c-ZrN and the hypothetical hexagonal w-ZrN structure as well as cubic and hexagonal structured $Zr_{0.5}Al_{0.5}N$ as ordered compounds. In the supercell systems, 64 atoms were included for c-$Zr_{0.5}Al_{0.5}N$, 108 atoms for w-ZrN, and w-AlN, while 128 atoms were included for c-ZrN, c-AlN, and w-$Zr_{0.5}Al_{0.5}N$.

## 3.2 Calculation of X-ray absorption and emission spectra

For the N *K*-edge XAS calculations, the core-hole approximation was used. In practice, an electron was removed from the N *1s* orbital at a specific nitrogen site and an extra electron added into the valence. This means that the charge neutrality of the system was kept constant. Here supercells of atoms were constructed in order to avoid self-interaction between core-ionized sites arising from periodic boundary conditions. The number of atoms in the supercells for XAS was 180 for the c/c, 160 for c/w, and 220 for c/w* structures. In addition, an 128 atoms $Zr_{0.75}Al_{0.25}N$ special quasi-random structure (SQS) [33] supercell with Zr and Al substitutional disorder was also studied.

For calculating the XES spectra, the so-called final-state rule [34] was applied using the XSPEC module in WIEN2k. Here no core-hole is needed and it is therefore sufficient to use unit cells in a regular ground state calculation. For both XAS and XES, the calculations were made within the electric-dipole approximation, meaning that only transitions to $l+/-1$ orbital angular momentum final states are included. Note that the spectral edges are computed separately for each inequivalent atomic site, thereafter the final spectrum for a structure is taken as a sum of the different spectra.

# 4 Results

## 4.1 Zr $M_{4,5}$ X-ray emission

Figure 3 shows Zr $M_{4,5}$ XES spectra of the $Zr_{1-x}Al_xN$ films (*x*=0, 0.12, 0.26 and 0.40) excited at 210 eV in comparison to a Zr metal reference film. The data are plotted on an emission energy scale (top) and a relative energy scale (bottom) to the $E_F$. The pure Zr material (black curves) was resonantly excited at the Zr $3d_{3/2}$ edge giving rise to elastic peaks at 180.0 and 182.4 eV. With excitation energy 210 eV, the pure Zr metal XES spectrum is dominated by a very intense peak with emission energy between 150-155 eV (25-30 eV below $E_F$) originating from Zr *4p->3d* dipole transitions from the shallow *4p$_{3/2}$* and *4p$_{1/2}$* core levels with 27.1 eV and 28.5 eV, respectively. This intense peak consists of two sub-peaks due to inner-transitions from the *3d$_{5/2,3/2}$* and *4p$_{3/2,1/2}$* spin-orbit splitting of 2.3 eV and 1.4 eV, respectively. The doublets can be observed in other parts of the spectra depending on the excitation energy and the Coster-Kronig process. The top of the valence band is observed in the energy region 173-180 eV due to Zr *4f5p->3d$_{5/2,3/2}$* dipole transitions while Zr *4d->3d$_{5/2,3/2}$* transitions are dipole forbidden.

The modeled Zr $M_{4,5}$ XES spectra including Zr *4p -> 3d* dipole transitions are shown below the measured data. The modelled crystal structures include cubic and hypothetical hexagonal w-ZrN (*x*=0) and the ordered compound Zr$_{1-x}$Al$_x$N (*x*=0.5). As observed in Fig. 3, the valence band edge shifts to higher energy with increasing *x*. The hybridization at the top of the valence band shows similarities with XES spectra of other nitrides *e.g.*, Ti$_2$AlN [35] and Sc$_3$AlN [36].

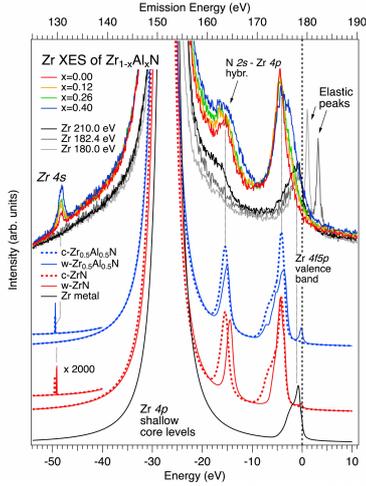

**Figure 3:** Measured (top) Zr $M_{4,5}$ XES spectra of Zr$_{1-x}$Al$_x$N for *x*=0.00, 0.12, 0.26 and 0.40. Bottom: Calculated cubic and hypothetical hexagonal w-ZrN (*x*=0) and Zr$_{1-x}$Al$_x$N (*x*=0.5) ordered compounds as model structures.

At the bottom of the valence band at 162-165 eV (16-17 eV from the $E_F$), an additional bond structure is observed, which is related to Zr *4p* - N *2s* hybridization. For this peak, the spectral weight shifts to towards the $E_F$ due to a decreasing *3d$_{5/2}$*/*3d$_{3/2}$* branching ratio as a function of Al content, which indicates decreasing conductivity with increasing Al content. Below the valence band and the intense *4p* peak, a small sharp peak located at 130 eV emission energy (50 eV below the $E_F$), increases in intensity with increasing Al content. The emission at 50 eV below $E_F$ is due to transitions between *4s* - *4p* hybridized Zr states to the *3d* level, where this emission gains dipole-intensity due to additional distortion with increasing *x* and this feature is completely absent in pure Zr.

The calculated spectra of the four candidate model systems at the bottom of Fig. 3 are in good agreement with the experimental results, except for the low-intensity peak at 50 eV below $E_F$ that is scaled up by a factor of 2000 (as indicated in Fig. 3) to be comparable in intensity with the experimental spectra. This scaling is necessary due to insufficient hybridization between the Zr *4s* – N *2s* orbitals at the bottom of the valence band that is inherent in the density functional theory (DFT). As similar bond situation with strong interaction at the bottom of the valence band has previously been observed in XES for shallow core levels of Ge [37] and Ga [15].

The spectra of pure Zr have a different spectral distribution than ZrAlN as there is no hybridization with N *2p* and N *2s* orbitals at 15-17 eV and 50 eV. In addition, the Zr *4p* states at the top of the valence band of pure Zr have less intensity and are located closer to the $E_F$ compared to ZrAlN. The spectral shapes of the modeled spectra are fairly similar, but for the hexagonal structure there is a clear energy shift of 1 eV towards the $E_F$ compared to the cubic systems. For ZrN, the modeled c-ZrN spectrum is in good agreement with the experimental ZrN data. On the other hand, for *x*=0.40, the modeled hexagonal type of Zr$_{0.5}$Al$_{0.5}$N structure has lower intensity and a shift towards $E_F$ of the Zr *4f5p* states of the valence band that is consistent with the observations in the experimental spectra. In order to find a more decisive difference between the cubic and hexagonal structures, we next examine the N *K*-edge.

Figure 4 shows measured N *K* XES spectra (top) of Zr$_{1-x}$Al$_x$N excited at 430 eV in comparison to *w*-AlN [38]. Calculated candidate spectra of cubic and hexagonal structures including the N *2p -> 1s* dipole transitions are shown at the bottom of Fig. 4. The main peak at 4-5 eV below $E_f$ mainly consists of bonding N *2p$_{xy}$*-σ states mixed with N *2p$_z$*-π states that are hybridized with the bonding Zr *4f5p* states. As observed, the N *K* XES spectrum of c-ZrN (*x*=0) has a flat high-energy tail reaching all the way up to the $E_F$, while the sharper main peak of *w*-AlN has a prominent low-energy structure at -8.5 eV below the top of the valence band edge.

Looking in detail at the energy position of the main peaks of the measured N *K* XES spectra of $Zr_{1-x}Al_xN$, we observe that there is a gradual shift of intensity of the main peak towards $E_F$ and a peak broadening with increasing *x*. In addition, intensity is redistributed from the high-energy tail at 0-3 eV to the low-energy shoulder at 7-10 eV below $E_F$. For *x*=0.12, this high-energy shift is negligibly small but the broadening is clearly noticeable in comparison to c-ZrN. For *x*=0.26, the high-energy shift is 0.15 eV and for *x*=0.40, the shift is increased to 0.7 eV that resembles the peak position of w-AlN.

Turning to the modeled N *K* XES spectra, the cubic spectrum of ZrN is in excellent agreement with the measured spectrum of ZrN. For comparison, the hypothetical w-ZrN has a 0.35 eV high-energy shift of the main peak towards the $E_F$ as well as a distinguished feature close to the $E_F$ that is not observed experimentally. Furthermore, the shape of the calculated w-AlN spectrum is in good agreement with the measured w-AlN spectrum including the position of the main peak and the low-energy feature. On the contrary, c-AlN has a different peak position and a low-energy feature that is more separated from the main peak. Based on this comparison, it can be concluded that c-ZrN is the dominant structure in the samples for low Al content, while for higher Al content, the ZrN contribution has a more hexagonal-like type of bonding with superimposed w-AlN, in particular for *x*=0.40.

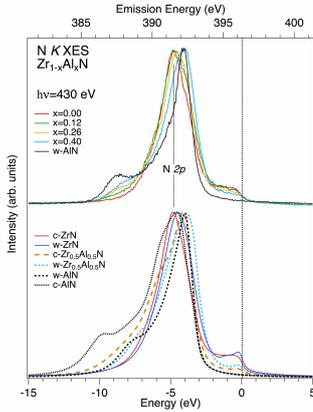

**Figure 4**, top: N *K* X-ray emission spectra of $Zr_{1-x}Al_xN$ in comparison to w-AlN. Bottom: calculated XES spectra of cubic and hexagonal ZrN, AlN and ordered $Zr_{0.5}Al_{0.5}N$ compounds.

The modeling shows that for the c-$Zr_{0.5}Al_{0.5}N$ structure, there is a redistribution of intensity from 0-3 eV below $E_F$ towards the low-energy shoulder at 7-10 eV with an additional shoulder at -6 eV on the side of the main peak. A similar intensity redistribution is observed for w-$Zr_{0.5}Al_{0.5}N$, but the decisive shoulder at the side of the main peak is absent as observed experimentally. For *x*=0.40, the w-$Zr_{0.5}Al_{0.5}N$ structure is thus in agreement with the experiment than c-$Zr_{0.5}Al_{0.5}N$. From the N *K* XES data, we thus conclude that the bonding structure around nitrogen gradually transforms and distorts from cubic to hexagonal in the nanocomposite and the spectral intensity represents a superposition of the contributions from two phases, ZrN and AlN.

Figure 5 (top) shows measured N *1s* XAS spectra of the unoccupied $2p_{xy}$ ($\sigma^*$) and $2p_z$ ($\pi^*$) orbitals probed via the N *1s* -> *2p* dipole transitions for *x*=0.00, 0.12, 0.26 and 0.40. The modelled N *1s* XAS spectra below the experimental data includes N *1s* -> *2p* dipole transitions for ideal ordered compounds with cubic and hexagonal structures w-ZrN (*x*=0) and $Zr_{1-x}Al_xN$ (*x*=0.5) as candidate structures. The modelling shows how the hexagonal distortion of the cubic structure affects the spectral features from alternating $t_{2g}^*$ and $e_g^*$ states in the cubic structure towards $\pi^*$ and $\sigma^*$ states in the hexagonal structure.

The peaks labeled A-H in Fig. 5 follow different trends. Starting with ZrN (*x*=0) at the bottom of Fig. 5, the agreement between experiment and theory for the prominent peaks B, C, D, F, G and H is very good for the cubic structure. In the cubic octahedral symmetry, the crystal-field of the Zr *4d* orbitals are divided into $t_{2g}$ and $e_g$ states containing the $d_{xy}$, $d_{xz,yz}$, and $d_{z2}$, $d_{x2-y2}$ orbitals, respectively. Cubic systems containing alternating metal-nonmetal elements are known to have metallic states with $t_{2g}$ symmetry ($\pi$-bonding) between the metal atoms and covalent (ligand-field) bonding with $e_g$ symmetry ($\sigma$-bonding) between the metal and non-metal atoms [13].

As the Zr *4d* orbitals in ZrN are surrounded by six N anion neighbors in the octahedral crystal field, these orbitals split into $t_{2g}$ and $e_g$ groups. The bonding and antibonding states are mainly due to the $e_g$ orbitals while orbitals of $t_{2g}$ symmetry do not significantly contribute to the covalent bonding. As a result of bonding with N (and Al in ZrAlN) the Zr $4d$-$e_g$ orbitals shift downward below the $E_F$, whereas the $t_{2g}$ orbitals shift upward above the $E_F$ due to Zr-Zr repulsion interaction. The result of the $t_{2g}$-$e_g$ splitting is the formation of a pseudogap between orbitals of bonding $e_g$ and nonbonding orbitals of $t_{2g}$ symmetry. The width of the pseudogap indicates the strength of the covalent bond that is a measure of the crystal stability that is thus related to the orbital hybridization.

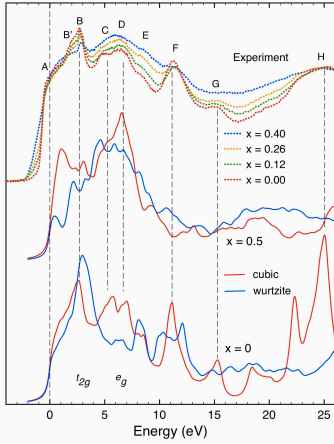

**Figure 5:** Experimental N *1s* XAS spectra (top) and calculated spectra (bottom) for cubic and hexagonal ZrN ($x=0$) and ordered $Zr_{0.5}Al_{0.5}N$ compounds.

For c-ZrN, the region 0-4 eV above the $E_F$ is dominated by peaks A and B where the unoccupied orbitals have metallic $t_{2g}$* character while in the energy region 4.5-9 eV, peaks C and D have covalent anti-bonding $e_g$* character. The shape of the XAS spectra of c-ZrN differ significantly from w-AlN that has a characteristic three-peak structure with three tetrahedrally coordinated σ* orbitals in the $2p_{xy}$ plane and a single π*-$2p_z$ orbital along the *c*-axis [39], but show similarities with cubic c-AlN with a single main N *1s* XAS peak [40]. For comparison to c-ZrN, the hypothetical spectrum of hexagonal w-ZrN structure in Fig. 5 differs significantly from the experimental spectra with peak structures at other energies above B and E and this structure can therefore be excluded.

The calculated spectrum of the cubic $Zr_{0.5}Al_{0.5}N$ ordered compound is dominated by peak D. The pre-peak shoulder A is also much more prominent in c-$Zr_{0.5}Al_{0.5}N$ than in c-ZrN that should influence the conductivity positively. Experimentally, the pre-peak feature A, increases in intensity and broadens with increasing Al content. The increasing intensity of the pre-peak A is a signature of emptier N *2p*\* states above $E_F$ that affects the conductivity of the system. Experimentally, the main peak denoted B in the XAS spectrum has the highest intensity in ZrN and has a dispersive shift away from the $E_F$ accompanied by a decreasing intensity for increasing Al content possibly associated with an increasing interface contribution or roughness. Another intensity trend is that of peak F that significantly decreases with increasing Al content while peak G is smeared out. The broad multi-electron structure denoted H, remains essentially the same, independent of Al content.

As evident from the calculations, the small and systematic high-energy shift of peak B due to $t_{2g}$ states observed experimentally is due to a gradual hexagonal distortion of the octahedral symmetry in the orbitals of the cubic structure. Peak B' is a low-energy shoulder of the main peak B that also decreases in intensity for increasing Al content. The double feature denoted by peaks C and D corresponding to $e_g$* states is most prominent in the cubic ZrN structure but merges and dissolves into one peak structure with higher intensity as the Al content increases.

Figure 6 shows calculated distributions of Zr *4d* states of $t_{2g}$ and $e_g$ symmetries around $E_F$ in the cubic c-ZrN and c-$Zr_{0.5}Al_{0.5}N$ structures in comparison to the hexagonal w-ZrN and w-$Zr_{0.5}Al_{0.5}N$ structures. The horizontal arrows in Fig. 6 indicate the pseudogaps between bonding Zr *4d* orbitals of $e_g$ symmetry and non-bonding Zr *4d* orbitals of $t_{2g}$ symmetry. The energy positions of the $e_g$-states change with structure and Al content as a result of the orbital distortion, redistribution (charge-transfer) and shift of intensity towards $E_F$. For w-ZrN, the pseudopap is slightly larger (8.93 eV) than for c-AlN (8.28 eV). On the contrary, c-$Zr_{0.5}Al_{0.5}N$ has a pseudogap of 7.62 eV compared to 7.08 eV for w-$Zr_{0.5}Al_{0.5}N$, indicating stronger bonding.

Changes in the orbital overlaps in terms of distance from the $E_F$ and integrated area of the $e_g$ peaks indicate changes in the bond strength as there is a shift towards $E_F$ and higher intensity in the unoccupied orbitals. With increasing Al content, the $e_g$ orbitals disperse towards $E_F$ and the resulting covalent Zr *4d*-$e_g$ – N *$2p_{xy}$*-σ bonding is weakened. The distortion and changes in bond strengths is also observed by the changes in Zr-N average bond lengths from 2.298 Å in c-ZrN to 2.226 Å in w-ZrN. However, the significantly increased number of states at $E_F$ for the w-ZrN structure in comparison to the stable c-ZrN structure makes the hexagonal structure unstable. From Fig. 6 it is clear that the covalent Zr *4d*-$e_g$ – N *$2p_{xy}$*-σ bonding weakens due to the shift of the $e_g$-$e_g$* orbitals towards the $E_F$ from both the occupied and unoccupied sides.

A complicating factor in the interpretation of the N *1s* XAS and N *K* XES data is the fact that the spectra contain a superposition of two different N-containing structures in addition to the interfaces. Moreover, for the modeled $Zr_{0.5}Al_{0.5}N$ ordered compound, the N-atoms coordinate at two different sites, one site that has a higher concentration of Al-nearest neighbors than the other site that has more Zr neighbors. Since a significant contribution to the N spectra originates from internal interfaces, we next compare the XAS spectra with different layer-resolved interfaces.

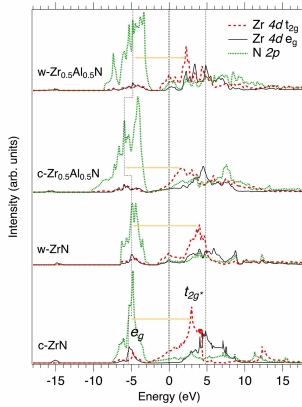

**Figure 6:** Calculated partial N *2p* and Zr *4d* DOS separated into $t_{2g}$ and $e_g$ symmetries. The horizonal arrows indicate the pseudogap between bonding $e_g$ orbitals and nonbonding orbitals of $t_{2g}$ symmetry.

Figure 7 shows calculated layer-resolved N *1s* XAS spectra of the three different interface structures shown in Fig. 2. The spectra at the top of the figure are averages over the entire interface slab, while the other six sets of spectra represent individual atomic layers in different parts of the interface slabs. The vertical dashed lines indicate the most prominent peak features originating from ZrN and AlN. The "bulk" AlN spectra at the bottom of Fig. 6 are a few atomic layers away from the AlN interface and the spectra from the three models show prominent peak structures between 2-8 eV at different energy positions. Notably, the bandgap of AlN is smallest for the asymmetric c/w* interface slab. Looking at the ZrN bulk contribution, peaks B, F and H are the most prominent features that can also be identified towards the ZrN interface (see vertical dashed lines). Notably, the average spectrum of the c/w* interface at the top of Fig. 6 with the most prominent peak structures B, F and H show the best agreement with the measured absorption spectra at the top of Fig. 5. In addition, the intensity distribution of peaks C and D is in good agreement. For comparison, the c/c and c/w average interfaces have peak features at other energy positions. Based on these observations, we conclude that the c/w* interface slab model best represents the experimental data. However, the calculated total energy of the cubic/cubic interface slab is lowest (-8.61 Ry/atom), and somewhat higher for the asymmetric cubic/wurtzite interface (-8.55 Ry/atom) while it is highest for the cubic/wurtzite interface (c/w=-8.10 Ry/atom).

Figure 8 (top) shows Al $L_{2,3}$ XES valence band spectra of $Zr_{1-x}Al_xN$ measured with a photon energy of 110 eV in comparison to w-AlN [39] as a reference material. In this case, the occupied Al *3sp* states are probed via the Al *3sp* -> *2p_{3/2,1/2}* dipole transitions. As observed, the experimental spectra are dominated by two broad peaks located at -4 eV and -7 eV below the $E_F$. As in the case of the Zr valence band edge, the measured Al *3sp* valence band edge shifts to higher energy for increasing *x*, most notable for *x*=0.40 where the two main peaks at -3.8 eV and -7.7 eV shifts by +0.4 eV towards $E_F$ for *x*=0.40 in comparison to *x*=0.12.

Calculated model spectra are shown at the bottom of Fig. 8 including ordered c/w-$Zr_{0.5}Al_{0.5}N$ compounds and a randomly disordered $Zr_{0.75}Al_{0.25}N$ alloy system in comparison to binary c/w-AlN as reference systems. The spectral shape of the calculated w-AlN reference spectrum is consistent with the measured w-AlN spectrum although the peak splitting of 3.8 eV is 0.45 eV larger than in the experiment (3.35 eV) and the peak located at -7.7 eV has higher intensity than the peak located at -4 eV that is opposite to what is observed in the experiment. For comparison, the calculated spectrum of c-AlN has a significantly larger peak splitting of 5.25 eV and a very different spectral shape than w-AlN. This is also the case for the calculated c-$Zr_{0.5}Al_{0.5}N$ spectrum that has a peak splitting of 4.20 eV. This shows that the AlN phase is not cubic for any *x*. For comparison, the w-$Zr_{0.5}Al_{0.5}N$ ordered compound model has a smaller peak splitting of 3.44 eV and a spectral shape with peak intensities that are in good agreement with the experiment. However, there is an additional peak feature at -0.2 eV just below $E_F$, most prominent in the c-$Zr_{0.5}Al_{0.5}N$ ordered compound that also appears with lower intensity in w-$Zr_{0.5}Al_{0.5}N$. This feature is due to the implicit perfect atomic periodicity in these ordered compound models and arises from the periodic boundary conditions. Thus, this peak feature does not appear for a randomly disordered $Zr_{0.75}Al_{0.25}N$ SQS alloy as shown in Fig. 8. The absence of this feature in the experimental spectra shows that this type of long-range ordering does not exist in the $Zr_{1-x}Al_xN$ nanocomposites.

## 5. Discussion

The electronic structure investigations of $Zr_{1-x}Al_xN$ solid solution and ZrN-AlN nanocomposite samples using bulk sensitive X-ray spectroscopies reveal several interesting observations. From the Zr $M_{4,5}$ XES, we observe that the *4f5p* states of the valence band gradually broaden and shift from 5 eV to 4 eV below $E_F$ with increasing Al content in the solid-solution state. For high Al content (*x*=0.40), the intensity of the valence band becomes reduced and broadened. This is a signature of a transformation from cubic to a more open structure like the hexagonal that should affect the material's hardness, elasticity, conductivity and other properties. The chemical shifts of Zr states to higher energy with increasing *x* indicate changes in charge transfer from Zr to N and Al. The energy shift is accompanied by increasing Zr *4d* - N *2s* orbital hybridization observed at the bottom of the valence band 16-17 eV below $E_F$.

From the N *K* XES, we find that the average bond structure around nitrogen in the ZrAl$_{1-x}$Al$_x$N solid solution gradually transforms from cubic in ZrN to a bonding resembling that of a hexagonal structure in the nanocomposite for *x*=0.40. The energy shift of the main N *K* emission peak by 1 eV in Fig. 4 is consistent with the shift observed for the Zr *4f5p* bonding states in Fig. 3 representing a bonding structure around nitrogen that gradually transforms and distorts from cubic to hexagonal with increasing Al content.

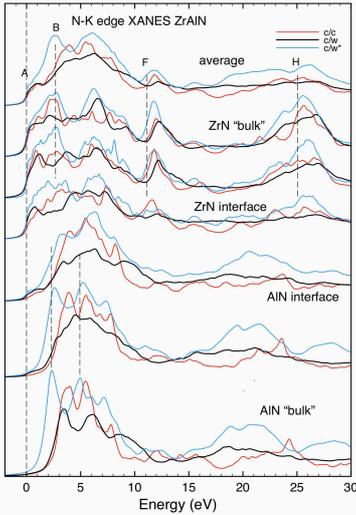

**Figure 7:** Calculated layer resolved N *1s* XAS spectra for the three different c/c, c/w and c/w* interface structures shown in Fig. 2. The spectra at the top are averages over the whole interface slabs.

In N *1s* XAS probing the unoccupied N *2p* states, we observe an intensity redistribution from the main peak associated with N *2p*-$t_{2g}$ states towards the N *2p*-$e_g$ states located at higher energy when the Al content increases. This is a signature of increasing emptier (broken) Zr *4d*-N *2p* covalent bonds of $e_g$ character and additional bonds with orbitals of other characters. The modelling shows how the $t_{2g}$* and $e_g$* peaks that are characteristic states of the cubic structure are affected by hexagonal distortion towards π* and σ* states in the hexagonal structure. A gradual smearing of the $t_{2g}$-$e_g$ splitting is also observed as the cubic crystal-field splitting diminishes with increasing Al content. At high Al content (*x*=0.40), a remaining weak peak-structure originating from cubic ZrN signifies a remaining superposition of some ZrN signal in the spectra while alternating $p_{xy}$-σ and $p_z$-π states resembles a hexagonal structure.

The modeled N *1s* XAS spectra of three different layer-resolved interface slabs show that at low Al-content (*x*=0.12), the c-ZrN/w-AlN model has the best fit with the experiment. At this early stage, ZrN and AlN precipitates appear to be coherent with no or little lattice mismatch between them. However, maintenance of coherency at higher Al contents becomes more demanding and implies lattice strain that affects dislocation movement and hardness of the material.

For higher Al content (*x*=0.26), an asymmetric c-ZrN/w-AlN* interface model is consistent with the experimental data. In this model, the lattice mismatch is 3.3% for c-ZrN/w-AlN interface that is larger than the mismatch of 2.2% for the c-ZrN/c-AlN interface. This gives rise to more compressive strain and less coherency (less lattice matching) for the former structure. The results are consistent with strain sensitivity studies using N *1s* XAS on GaN [41] that showed enhanced intensity of the main peaks with increasing interface strain. At high Al content (*x*=0.40), diffusion causes transformation from Zr$_{1-x}$Al$_x$N solid solution into ZrN and AlN precipitates that likely increases the lattice mismatch giving rise to an evolution of properties such as the resistivity and hardness.

From Al $L_{2,3}$ XES, we observe that the spectral intensity distribution of AlN largely resembles a model of an ordered w-Zr$_{0.5}$Al$_{0.5}$N alloy and is not cubic. If Al has segregated and formed AlN-rich domains, such domains has a hexagonal structure and there is no indication of pure Al domains. A predicted emission peak near the $E_F$ does not appear for randomly oriented alloys but only for highly ordered compounds.

Table I shows calculated Bader charges where more charge (1.8960e) is transferred from Zr to N for cubic ZrN compared to w-ZrN (1.8662e), indicating stronger Zr-N bonding in the cubic case. This is also the case for c-Zr$_{0.5}$Al$_{0.5}$N but additional charge is transferred from Al to N and more charge is transferred from Al in the hexagonal system. The additional charge of 0.1e from Al indicates that the Al-N bonding is stronger in the hexagonal system.

**Table I:** Calculated Bader charges of cubic and hexagonal ZrN and $Zr_{0.5}Al_{0.5}N$.

| System | Zr | Al | N |
|---|---|---|---|
| c-ZrN | 1.8960 | - | -1.8960 |
| w-ZrN | 1.8662 | - | -1.8662 |
| c-ZrAlN | 1.9763 | 2.2693 | -2.1228 |
| w-ZrAlN | 1.8587 | 2.3681 | -2.1134 |
| c-AlN | - | 2.4378 | -2.4378 |
| w-AlN | - | 2.3780 | -2.3780 |

Previous TEM studies on $Zr_{1-x}Al_xN$ films grown under similar conditions as the ones studied here showed [10] [11] that these films did not consist of pure solid solution alloys but of nanostructures of AlN embedded in ZrN-rich $Zr_{1-x}Al_xN$ matrix. This is consistent with the XRD results in Fig. 1. For $x$=0.26, the ZrN 400 XRD peak in Fig. 1 exhibits a shift towards lower 2θ angles indicating tensile strain of the ZrN lattice. On the other hand, for $x$=0.40, the 400 peak shift indicates compressive strain in ZrN. This may be associated with the formation of w-AlN inclusions and the smaller atomic size of Al. The strain also has an effect on the X-ray spectra although it probes short-range order while XRD probes long-range order.

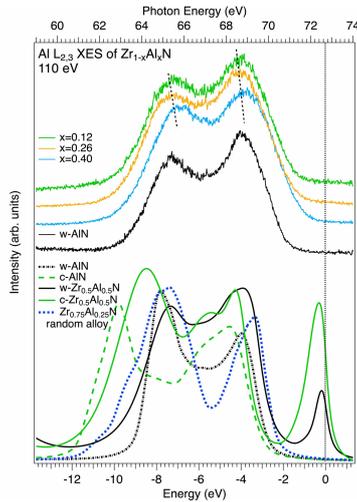

**Figure 8:** Measured Al $L_{2,3}$ XES spectra (top) of $Zr_{1-x}Al_xN$ and w-AlN in comparison to calculated spectra using different models (bottom).

Our calculations of the density of states (DOS) for the structured $Zr_{0.5}Al_{0.5}N$ compounds show higher number of states at the $E_F$ for c-$Zr_{0.5}Al_{0.5}N$ (0.96 states/eV/atom) compared to c-ZrN (0.69 states/eV/atom). According to *Mott's law of conductivity* [42] [43], the higher number of states at the $E_F$ for c-$Zr_{0.5}Al_{0.5}N$ versus c-ZrN should increase the conductivity due to the metal bonding and affect the stability by charge redistribution in the valence band. This is also the case for the hexagonal w-$Zr_{0.5}Al_{0.5}N$ structure (3.9 states/eV/atom) compared to w-ZrN (2.5 states/eV/atom). However, this rough approximation is opposite to experimental resistivity measurements that increases with Al content from 13.6 μΩ·cm in ZrN, to 200 μΩ·cm at $x$=0.3 and 15000 μΩ·cm at $x$=0.8 [44] and exponentially with temperature [45].

The increasing resistivity with increasing Al content is observed in the Zr $M_{4,5}$ XES as a change of spectral weight that shifts to towards the $E_F$ due to a decreasing $3d_{5/2}/3d_{3/2}$ branching ratio as a function of Al content. It is also observed by the increasing intensity of the pre-peak in N $1s$ XAS is more prominent in c-$Zr_{0.5}Al_{0.5}N$ than in c-ZrN that is a signature of emptier N $2p$* states above $E_F$ that increases the resistivity. The increasing chemical shift and charge transfer between the elements takes place with increasing Al content that increases resistivity and affects the bond strength in the structure. The transformation from a pure cubic structure to a more open structure resembling to that of the hexagonal also affect the material's conductivity and other transport properties. At high Al content, diffusion from $Zr_{1-x}Al_xN$ solid solution decompose the $Zr_{1-x}Al_xN$ solid solution into ZrN and AlN precipitates that likely increases the lattice mismatch and strain between nanosized lamella likely increases the resistivity as well as hardness.

Concerning the film hardness behavior, our present spectroscopic results provide support for earlier assumptions made for the phase composition and hardening mechanisms [11]. It was shown [11] that $Zr_{1-x}Al_xN$ films exhibits an increasing hardness with increasing Al content reaching a maximum for $x$=0.36 from where it decreases. The increased hardness with increasing Al content was suggested to be due to an increasing amount of strain at the interfaces between the ZrN and the AlN precipitates, which we have identified in the N $1s$ XAS spectra in Fig. 5 as a gradual shift of the N $2p$-$e_g$* states away from the $E_F$ supported by the calculations. Experiments have shown that the Young's modulus increases from 250 GPa in c-ZrN to 300 GPa in $Zr_{0.57}Al_{0.43}N$ consistent with calculations [45] as well as XRD of ZrAlN multilayers [46]. Furthermore, wide-angle X-ray scattering [47] has shown that the ZrAlN system exhibit high thermal phase stability [48] [49] and hardness [50] during annealing. This is in accordance with the thermal stability observed here. Increasing amount of interfaces result in high hardness and the highest hardness is found when the interfaces are semi-coherent, which is also supported by the present N *1s* XAS spectra in comparison to the layer-resolved calculations shown in Fig. 7. However, maintenance of coherency at higher Al contents becomes increasingly demanding due to the differences in atomic size and implies lattice strain that affects dislocation movement and an evolution of the resistivity, elasticity and hardness properties of the material. An increase of the interface contribution is observed both in the *4f5p* states in the Zr $M_{4,5}$ and the N *2p* states in the N *K* XES as a gradual shift and a broadening of the upper part of the valence band associated with the covalent Zr-N bonding. The broadening of the valence band associated with orbital overlaps implies less directional Zr-N bonds as the Al content increases.

In contrast to the changes in the valence band, sharp hybrid Zr-N states are observed at a constant energy of 50 eV below $E_F$ for ZrN with an increasing intensity as a function of Al content. At sufficiently high Al content, diffusion during the growth occurs and would further decompose and segregate the $Zr_{1-x}Al_xN$ solid solution into c-ZrN and w-AlN precipitates, as supported herein both by the Al $L_{2,3}$ XES spectra in Fig. 8 that largely resemble the spectral shape of w-AlN for high Al content and not c-AlN. With increasing amount of wide-band-gap AlN phase and interface contribution, the film resistivity should also increase [44]. This trend is observed both as a decreasing amount of N *2p* states in the tail at the top of the valence band in the N *K* XES spectra in Fig. 4 as well as an increasing amount of unoccupied N *2p\** states in the pre-edge shoulder of the N *1s* XAS spectra in Fig. 5 as a function of Al content. However, as shown by the calculated interface slabs in Fig. 7, if the AlN lamellar precipitates consists of a few atomic layers only, interface states appear in the atomic layers close to the interface while a band gap only appears in the bulk of the material. Thus, understanding the interface bonding in these kinds of nanocomposites enables design of the materials properties.

## 6. Conclusions

We have investigated the local electronic structure and interface bonding using soft X-ray absorption and resonant inelastic X-ray scattering/X-ray emission measurements on the metastable $Zr_{1-x}Al_xN$ compound in comparison to the corresponding binaries c-ZrN and w-AlN. The experimental spectra are compared with different interface-slab model systems using first-principle all electron full-potential calculations where the core states are treated fully relativistic. We show how the electronic structure develop from local octahedral bond symmetry of cubic ZrN that distorts for increasing Al content into more complex bonding. The distortion results in three different kinds of bonding originating from semi-coherent interfaces with nanocrystalline segregated lamellar AlN precipitates. For low to moderate Al content ($x$=0.12, 0.26), our model calculations show that the interface is c-ZrN/w-AlN that is consistent with the experimental data. At this Al content, diffusion likely results into c-ZrN and w-AlN segregated phases that has been referred to as a nanolabyrinth containing nanolamellar structure as previously observed in TEM. As the Al content increases, the $t_{2g}$-$e_g$ crystal-field splitting characteristic of a cubic ZrN system is gradually diminished and a significant change of bond character towards a hexagonal structure is observed. At $x$=0.40, the shifts of the valence bands as well as the changes in the intensities of the Zr, N, and Al peak structures show that the ZrN structure is highly distorted with a more open crystal structure than that of the cubic. Comparison to calculated interface model structures confirms that the $Zr_{1-x}Al_xN$ nanocomposite gradually transforms into a system with asymmetric ZrN/wurtzite-AlN interfaces that largely resembles that of a hexagonal $Zr_{1-x}Al_xN$ compound. The sensitivity of X-ray spectroscopy allows to distinguish between the orbital occupations of the internal interfaces that determines the bond characteristics of cubic and hexagonal structures and local interface chemistry that affects conductivity, elasticity and hardness.

### 7. Acknowledgements
We thank the Swedish Research Council (VR) LiLi-NFM Linnaeus Environment and project Grant No. 621-2009-5258, the staff at the MAX IV laboratory for experimental support. The research leading to these results has received funding from the Swedish Government Strategic Research Area in Materials Science on Functional Materials at Linköping University (Faculty Grant SFO-Mat-LiU No. 2009-00971). M.M. acknowledges financial support from the Swedish Energy Research (no. 43606-1) and the Carl Tryggers Foundation (CTS16:303, CTS14:310). W.O. acknowledge the Swedish Government Strategic Research Area in Materials Science on Functional Materials at Linköping University (Faculty Grant SFO-Mat-LiU No 2009 00971) and Knut and Alice Wallenberg Foundation project Strong Field Physics and New States of Matter CoTXS (2014–2019). We thank Dr. Björn Alling for providing the SQS $Zr_{0.75}Al_{0.25}N$ random alloy structure.